\documentclass[singlecolumn,superscriptaddress,aps,showpacs,floatfix,prb,12pt]{revtex4-1}
\usepackage[T1]{fontenc}
\usepackage{amssymb,amsmath}
\usepackage{graphicx}
\usepackage{color}
\usepackage{bbold}
\usepackage{soul}
\usepackage[dvipsnames]{xcolor}
\usepackage{ulem}
\newcommand{\ah}{\hat{a}}
\newcommand{\bh}{\hat{b}}

\begin{document}
\title{Single photons from a gain medium below threshold}
\author{Sanjib Ghosh}
\email{e-mail: sanjibghosh87@gmail.com}
\affiliation{Division of Physics and Applied Physics, School of Physical and Mathematical Sciences, Nanyang Technological University, 21 Nanyang Link, Singapore 637371}
\author{Timothy C. H. Liew}
\email{e-mail: tchliew@gmail.com}
\affiliation{Division of Physics and Applied Physics, School of Physical and Mathematical Sciences, Nanyang Technological University, 21 Nanyang Link, Singapore 637371}

\begin{abstract}
The emission from a nonlinear photonic mode coupled weakly to a gain medium operating below threshold is predicted to exhibit antibunching. In the steady state regime, analytical solutions for the relevant observable quantities are found in accurate agreement with exact numerical results. Under pulsed excitation, the unequal time second order correlation function demonstrates the triggered probabilistic generation of single photons well separated in time.
\end{abstract}


\maketitle

{\bf Introduction.---} Single photon sources are an essential component for emerging quantum technologies such as quantum computation~\cite{Knill01}, quantum cryptography~\cite{Scarani09} and long distance quantum communications~\cite{Sangouard11,Kimble08}. Pulses from a faint laser are often taken to constitute a single photon source, however, even the faintest laser generates multiphoton pulses as the photon number obeys Poissonian statistics. These unavoidable multiphoton pulses are unsuitable for many applications~\cite{Brassard00}. This motivates the study of quantum nonlinear systems, where Poissonian statistics can be skewed to favour antibunched light sources.

Mechanisms of generating antibunched light typically rely on coherent resonant excitation. To give examples, parametric down conversion requires phase matching conditions to be achieved and the photon blockade mechanism~\cite{Imamoglu97} is based on the interplay of an anharmonic energy spectrum with the specific frequency of a coherent source~\cite{Birnbaum05,Dayan08,Faraon08,Reinhard11,Lang11}. An alternative blockade mechanism known as the unconventional blockade~\cite{Liew10,Bamba11,Lemonde14,Flayac17} has been recently reported experimentally~\cite{Vaneph18} using superconducting resonators. It illustrates that the quantum optics of two coupled nanophotonic modes can be vastly different to that of a single mode~\cite{Carmichael85} and that the range of open quantum systems for observing quantum optical effects is steadily increasing. At the same time, it illustrates further the tendency of open quantum systems to operate with coherent sources when the objective is a non-classical state.

Photonic resonators containing a gain medium are also well studied, where gain represents excitation through scattering processes that are not themselves coherent. It is only above threshold that the scattering processes become stimulated and allow the formation of a coherent state, characterized by Poissonian statistics. Below threshold, a single resonator exhibits an incoherent state of small bunched number fluctuations. In either regime, the gain medium does not seem particularly well suited to observing antibunched states.

Here, we recall that the physics of coupled quantum modes may be different. We consider a generic open quantum system comprised of a strongly nonlinear mode weakly coupled to a gain medium operating below the single-mode threshold. Presenting analytic and numerical solutions for the master equation in the steady state, we show that photons passing from the gain medium to the nonlinear mode can generate an antibunched state. At the same time, the mean-field occupation of the modes remains zero and the state of the gain medium remains incoherent, representing a situation radically different to that of existing blockade mechanisms.

We identify a pair of coupled exciton-polariton modes in semiconductor microcavities as an example of a potential physical realization. Semiconductor microcavities are well-known for functioning as a gain medium where under electrical excitation they realize light-emitting diodes~\cite{Tsintzos08} and polariton lasers~\cite{Schneider13,Bhattacharya13,Zhang14}. Furthermore, polaritons are known to behave as quantum particles, passing their quantum properties into an emitted optical field~\cite{Cuevas16,Adiyatullin17} and their antibunching was experimentally reported under coherent excitation~\cite{Matutano17}. 
While only showing the weak onset of the polariton blockade~\cite{Verger06}, the strongly nonlinear regime (where the interaction strength between a pair of interacting polaritons exceeds their linewidth) has been reached in separate experiments~\cite{Sun17,Rosenberg18,Togan18}.

Finally we consider the situation of a pulsed excitation or time-dependent gain, where we find strong antibunching during time periods when the nonlinear mode is significantly populated. By calculating an unequal time correlation function, we show that for an appropriate choice of measurement time window single photons are generated at moments well separated in time. Thus the considered system is capable of triggering single photons with some probability.

{\bf Theoretical Scheme.---} We begin with the Hamiltonian describing two coupled bosonic quantum modes (a Bose-Hubbard dimer):
\begin{eqnarray}
\hat{H} = \epsilon_a\, \ah^\dagger \ah + \alpha\, \, \ah^\dagger \ah^\dagger \ah \ah +  \epsilon_b \,\bh^\dagger \bh+  J\, ( \ah^\dagger \bh+ \bh^\dagger \ah  )
\label{Hamiltonian}
\end{eqnarray}
where $\ah$ and $\bh$ are the annihilation operators; $\epsilon_a$ and $\epsilon_b$ are the respective uncoupled energies of the two modes. $J$ is the coupling strength between $\ah$ and $\bh$. The parameter $\alpha$ describes a Kerr-type nonlinearity of the $\ah$ mode, while the other $\bh$ mode is considered linear. The system could be physically realized with photonic crystal cavities~\cite{Gerace09,Ferretti13} superconducting circuits~\cite{Rosenberg18} or coupled micropillars~\cite{Michaelis11,Rodriguez18} containing exciton-polaritons. We note that in the latter system the coupling $J$ is controllable through the micropillar overlap and the micropillar size (which in principle could be different for the two micropillars) affects the effective nonlinear interaction strength $\alpha$ by changing the mode volume~\cite{Verger06}. Alternative methods of localizing exciton-polaritons into discrete modes are reviewed in Ref.~\cite{Fraser17}. Assuming that the mode $\bh$ corresponds to a gain medium (which in the case of micropillars corresponds to the non-resonant excitation of an exciton reservoir~\cite{Galbiati12} in the micropillar containing mode $\bh$), the system is described by the quantum master equation for the density matrix $\rho$:
\begin{eqnarray}
i\hbar \frac{d\rho}{dt} = [\hat{\mathcal{H}},\rho]  + \mathcal{L}(\rho) + \frac{i P_b}{2} \left( 2 \bh^\dagger \rho \bh  - \rho\bh  \bh^\dagger  - \bh  \bh^\dagger \rho \right)
\label{LindbladEquation}
\end{eqnarray}
where $ \mathcal{L}(\rho) = \sum_{s=a,b} i\gamma_s (  \hat{s} \rho  \hat{s}^\dagger  - \rho \hat{s}^\dagger  \hat{s}/2 - \hat{s}^\dagger \hat{s}  \rho/2 )$ is the Lindblad term describing dissipation in the two modes. While the first term on the right-hand-side of Eq.~\ref{LindbladEquation} represents the coherent evolution, the last term represents a gain applied to the linear mode $\bh$ which can be interpreted as a time reversed dissipation (such a form has appeared previously in the context of quantum dots \cite{Laussy08}). $\gamma_a/\hbar$ and $\gamma_b/\hbar$ are the dissipation rates of the nonlinear mode $\ah$ and the gain mode $\bh$ respectively, and $P_b/\hbar$ is the gain rate in mode $\bh$. All numerical data presented in this letter will be obtained by exact numerical simulations of Eq.~\ref{LindbladEquation} using a truncated Fock basis\cite{FootNote2}.

Although the gain is applied to the mode $\bh$, we will focus on the statistics of the mode $\ah$. As a measure of antibunching, we calculate the unequal time second order correlation function $g_2(t_1,t_2)$, defined by:
\begin{eqnarray}
g_{2}(t_1,t_2) = \frac{  \langle \,\ah^\dagger (t_1) \ah^\dagger (t_2) \ah(t_1) \ah(t_2) \rangle }{   \langle \,\ah^\dagger (t_1)\ah (t_1) \rangle  \, \,\,   \langle \ah^\dagger(t_2) \ah(t_2)  \rangle }
\label{CorrelationFunction}
\end{eqnarray}
where $ \langle \dots  \rangle$ denotes the expectation value of the respective operators. We recall that the equal time correlation function $g_2(t,t)$ evaluates to one for a coherent (classical) state and is zero for the ideal single-particle state.

\begin{figure}[h]
\includegraphics[width=1\columnwidth]{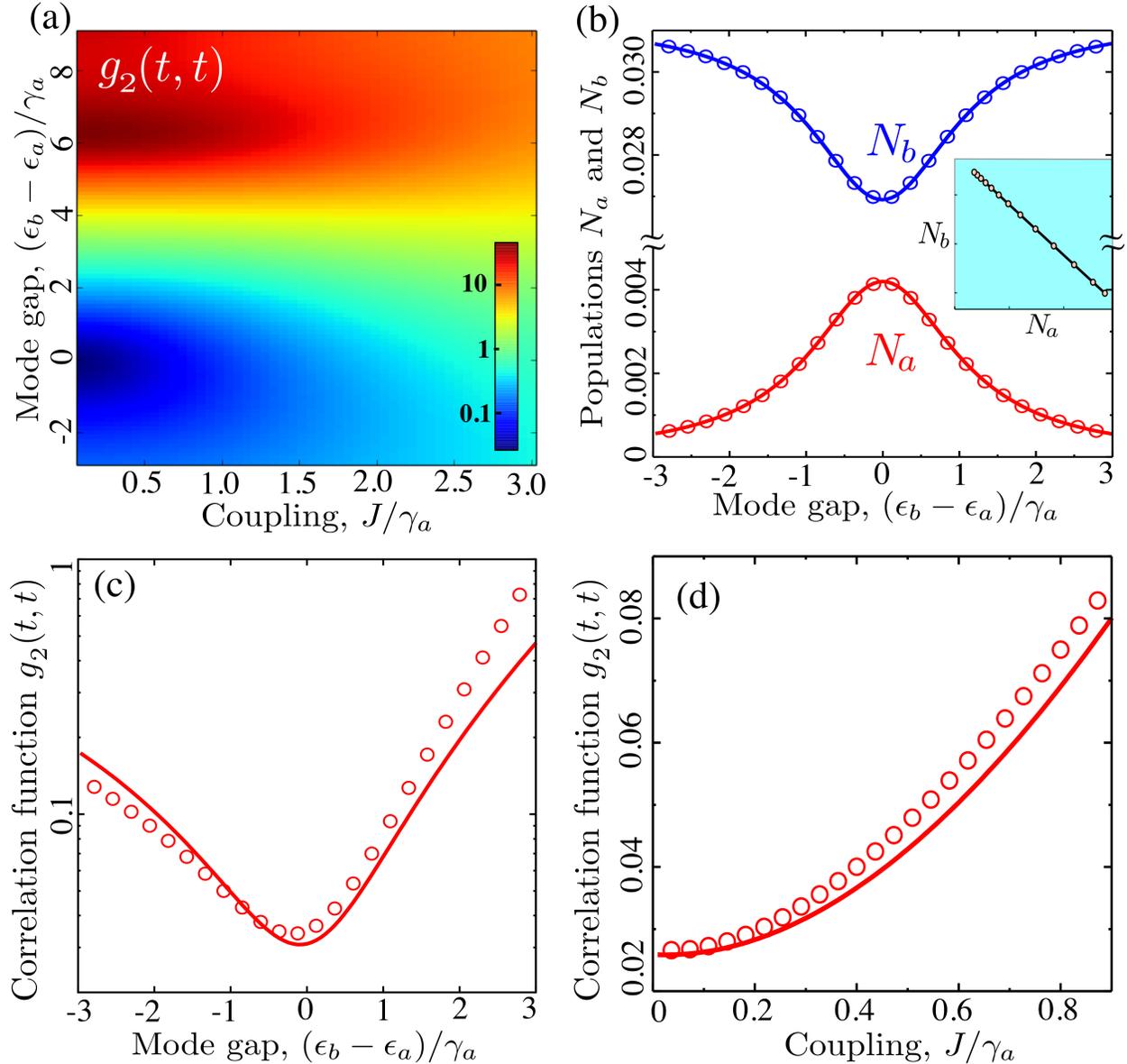}
\caption{The steady state properties of the system with a constant gain in mode $\bh$. (a) Color plot of the equal time correlation function $g_2(t,t)$ calculated for the nonlinear mode $\ah$ as a function of the mode energy gap $\epsilon_b-\epsilon_a$ and coupling strength $J$. A strong antibunching occurs in the weak coupling $J/\gamma_a \ll 1$ regime for $\epsilon_b-\epsilon_a \approx 0$ (deep blue region). (b) Mode populations $N_a = \langle \ah^\dagger \ah \rangle$ in mode $\ah$ and $N_b=\langle \bh^\dagger \bh \rangle$ in mode $\bh$ as function of $\epsilon_b-\epsilon_a$ calculated using the analytic method (solid lines) given by Eq.~\ref{InterrelationNaNb} and \ref{AnalyticC}; and by exact numerical simulations (open circles). While $N_b$ remains nonzero for all energies, $N_a$ becomes significant only for $\epsilon_b-\epsilon_a\approx 0$. In the inset, we replot the data (circles) as $N_a$ versus $N_b$ to show that they are linearly related, in agreement with Eq.~\ref{InterrelationNaNb} (solid line). (c) and (d) show $g_2(t,t)$ as functions of $\epsilon_b-\epsilon_a$ and $J$ respectively, calculated analytically using Eq.~\ref{AnalyticG2} (solid lines) and numerically (open circles). The data are obtained with parameters fixed at $\alpha/\gamma_a=6.06$, $(\epsilon_b-\epsilon_a)/\gamma_a=-0.303$, $J/\gamma_a=0.303$, $\gamma_b/\gamma_a=1$ and $P_b/\gamma_a=0.0303$ other than the running variables. We note that a similar ratio of $\alpha/\gamma_a$ was achieved in Ref.~\cite{Togan18} using an exciton-polariton system. For (d) we chose the optimum $\epsilon_b-\epsilon_a$ minimizing $g_2(t,t)$ for each value of $J$.}
\label{SteadyStateFig}
\end{figure}

For a constant gain $P_b$, the system reaches, as a generic feature, a steady state after some initial time evolution. In such a state, the mode $\ah$ loses particles at a constant rate. We calculate the equal time correlation function $g_2(t,t)$ using Eq.~\ref{CorrelationFunction} for this mode. In Fig.~\ref{SteadyStateFig} (a), we present $g_2(t,t)$ as a function of the mode coupling $J$ and the energy gap $\epsilon_b-\epsilon_a$.~We observe a strong antibunching effect ($g_2(t,t)\sim 0$) when $\epsilon_a\approx\epsilon_b$ and the mode coupling is weak, given by the blue area in the figure. The closing mode gap $\epsilon_b \approx\epsilon_a$ allows particles from the gain mode to efficiently transfer to the nonlinear mode. In this regime, we find a maximum population in the nonlinear mode, while a minimum population appears in the gain mode, see Fig.~\ref{SteadyStateFig}(b). However, nonlinearity suppresses multi-particle occupations and thus lowers $g_2(t,t)$ in the $\ah$ mode. One might hope to interpret this as the gain mode representing an effective coherent source that acts on the nonlinear mode in the same way as a laser in the case of the photon blockade. However, the mean field population of both modes, $\langle \ah \rangle$ and $\langle \bh \rangle$, vanishes and we verified that the $\bh$ mode is far from coherent (as we operate below threshold). Consequently, the physics is significantly different to previous examples of photon/polariton blockade.

{\bf Analytical Interpretation.---} To interpret the results, we can instead study the steady state solutions analytically. Writing equations of motion for $N_a=\langle \ah^\dagger \ah \rangle$ and $N_b=\langle \bh^\dagger \bh \rangle$ and then taking $N_a$ and $N_b$ as constant, we deduce that:
\begin{eqnarray}
N_b = (P_b-\gamma_a N_a )/( \gamma_b-P_b )
\label{InterrelationNaNb}
\end{eqnarray}
We focus on the below threshold regime, $P_b<\gamma_b$, where an increasing $N_a$ imposes a decrease in the steady state population $N_b$. This behavior is evident in Fig.~\ref{SteadyStateFig}(b) and the inset. However, Eq.~\ref{InterrelationNaNb} alone is not enough to find $N_b$ and $N_a$ individually, which require finding $C=\langle \ah^\dagger \bh\rangle$. A mere mean field approximation of type $C \approx \langle \ah^\dagger \rangle \langle \bh\rangle$ breaks down, since $\langle \ah \rangle=0$. The time evolution of $C$ can be obtained from the master equation and depends on second order correlations like $\langle \ah^\dagger \bh \ah^\dagger \ah \rangle$. It turns out that this second order correlation is crucial for accurate evaluations of $N_a$, $N_b$ and $g_2(t,t)$. Using the steady state equation for $\langle \ah^\dagger \bh \ah^\dagger \ah \rangle$ and approximating the further higher order correlations in terms of $C$, $N_a$ and $N_b$, we arrive at a solution valid for $J/\gamma_a \ll 1$ and $\alpha/\gamma_a\gg 1$:
\begin{eqnarray}
C  \approx  \frac{ J }{ E_1}(N_b-N_a) +  \frac{ 4J\alpha }{E_2  E_1}  \,N_a N_b
\label{AnalyticC}
\end{eqnarray}
where the energies are given by $ E_1 = (\epsilon_b-\epsilon_a)  - i( \gamma_a+\gamma_b-P_b )/2 $ and $ E_2 = (\epsilon_b-\epsilon_a) - 2\alpha - i(3\gamma_a+\gamma_b-P_b)/2$.
Note that the imaginary part of $C$, $\text{Im}C$, represents the current of population flow from mode $\bh$ to mode $\ah$. This current induces accumulation of population in mode $\ah$: $N_a = (2J/\gamma_a)\text{Im}C$ in the steady state. In fact, this relation and Eq.~\ref{InterrelationNaNb} together yield a quadratic equation: $N^2_a-2\zeta_1N_a+\zeta_2=0$ (equivalently for $N_b$) where coefficients $\zeta_1$ and $\zeta_2$ are solely given by the system parameters $\gamma_a$, $\gamma_b$, $P_b$, $J$ and $\alpha$. In Fig.~\ref{SteadyStateFig}(b), we compared this fully-analytic solution\cite{FootNote1} with the exact populations $N_a$ and $N_b$ numerically calculated using Eq.~\ref{LindbladEquation}. Despite all approximations made, the analytical solutions show excellent agreement with the numerical results as shown in Fig.~\ref{SteadyStateFig}(b). Beyond these single particle observable quantitites, we find an analytical solution for $g_2(t,t)$:
\begin{eqnarray}
g_2(t,t)  = \frac{J\,(\epsilon_b-\epsilon_a) }{\alpha \gamma_a N_a^2} \text{Im}C -\frac{J\,(\gamma_a+\gamma_b-P_b)}{2\alpha\gamma_a N_a^2} \text{Re}C
\label{AnalyticG2}
\end{eqnarray}
where $C$ is calculated from Eq.~\ref{AnalyticC} aided by the previously obtained formula for $N_a$ and $N_b$.

In Fig.~\ref{SteadyStateFig}(c) and (d), we compare $g_2(t,t)$ given by Eq.~\ref{AnalyticG2} to exact numerical results as functions of the mode gap $\epsilon_b-\epsilon_a$ and coupling strength $J$. We observe that the agreement between the analytical and numerical results is almost exact for small $J$. The reason can be traced back to Eq.~\ref{AnalyticC} which is found to be exact for $J/\gamma_a \ll 1$. Moreover, only a weak intermodal coupling induces strong single photon statistics (small $g_2(t,t)$) as evident in Fig.~~\ref{SteadyStateFig}(d). Thus, our analytical solution given in Eq.~\ref{AnalyticC} is nearly exact for the most relevant regime of the system. The effects that can skew the single photon statistics are a strong nonlinearity in the pumped mode $\bh$ or a weak nonlinearity in mode $\ah$. However, all these parameters can effectively be tuned in modern experimental setups.

\begin{figure}[h]
 \includegraphics[width=1\columnwidth]{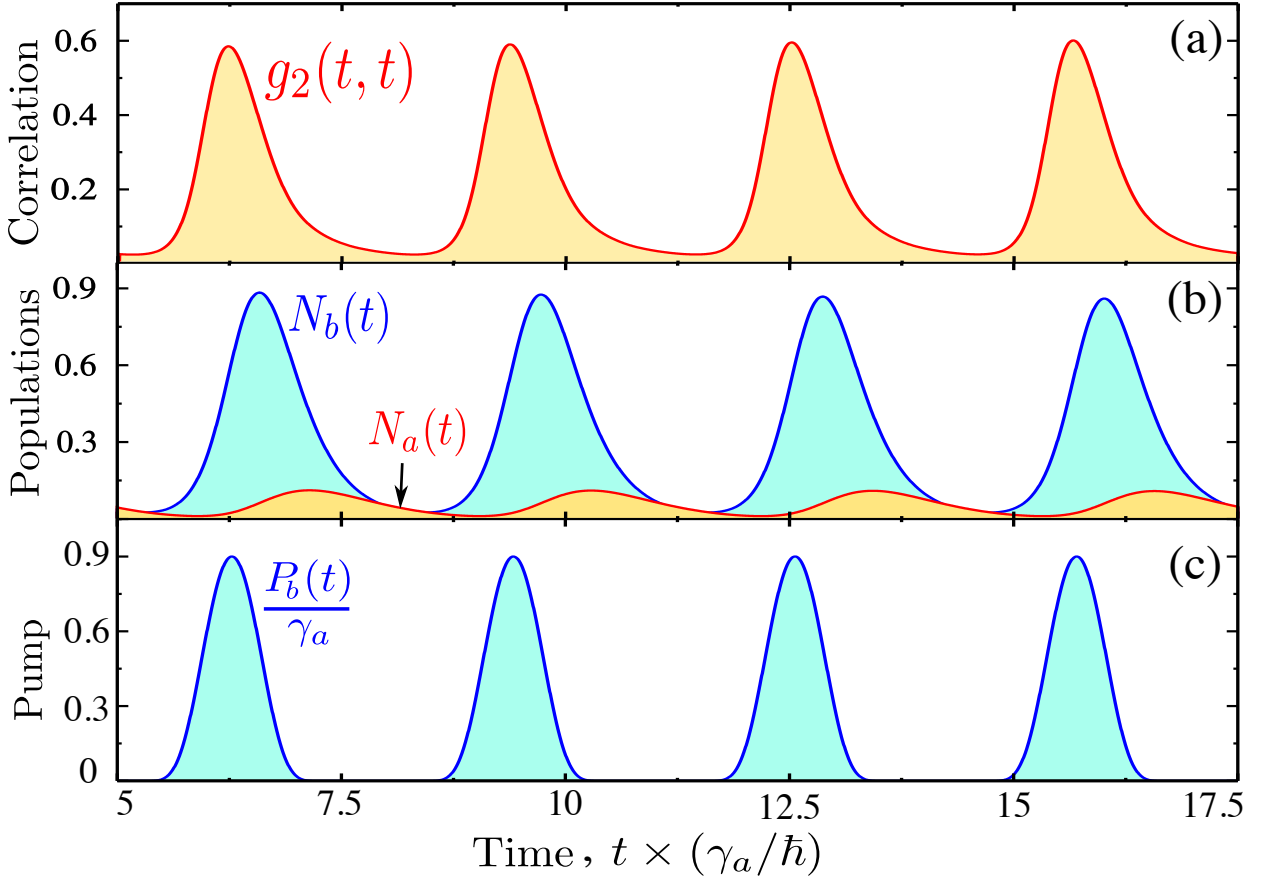}
 \caption{Time evolution of the system when excited by a series of gain pulses. (a) shows the equal time correlation function $g_2(t,t)$ and (b) shows the time-dependent populations $N_a(t)$ (red line) and $N_b(t)$ (blue line) as responses to the incoherent pulses $P_b(t)$ shown in (c) applied to the gain mode $\bh$. In each pulse, $g_2(t,t)$ decreases down to $\sim 0.03$ (strong single particle statistics), while $N_a$ rises up to $\sim 0.11$. The data are obtained with parameters $\alpha/\gamma_a=6.06$, $(\epsilon_b-\epsilon_a)/\gamma_a=-0.545$, $J/\gamma_a=0.76$ and $\gamma_b/\gamma_a=1$. Pulses are generated using a periodic function $P_b(t)=P_0 \exp[-A\sin^2(\pi t/T_0)] $ with $P_0/\gamma_a=0.91$, $T_0=\pi\,\hbar/\gamma_a$ and $A=5$.}
\label{PulseTrainFig}
\end{figure}

{\bf Pulsed Gain.---} We now consider the situation of time-varying gain, assuming that it is possible to engineer a series of gain-inducing pulses of the form $P_b(t)=P_0 \exp[-A\sin^2(\pi t/T_0)]$ that act on the mode $\bh$. We turn on the pump at $t=0$ with an initial condition $N_a=N_b=0$. Following some transient dynamics the observable quantities in the system like $g_2(t,t)$, $N_a(t)$ and $N_b(t)$ become periodic in time. In Fig.~\ref{PulseTrainFig}, we find that while the time modulation in $N_b(t)$, more or less, follows the pump $P_b(t)$, the modulation in $N_a(t)$ has a time delay. This delay can be associated to the time taken to transfer photons from the gain mode $\bh$ to the $\ah$ mode. Comparing Fig.~\ref{PulseTrainFig} (a) and (b), we find that the $\ah$ mode shows a very small $g_2(t,t)$ when population $N_a(t)$ is significant. Thus, even with the pulses, we have a significant antibunching statistics.

\begin{figure}[h]
\includegraphics[width=1.0\columnwidth]{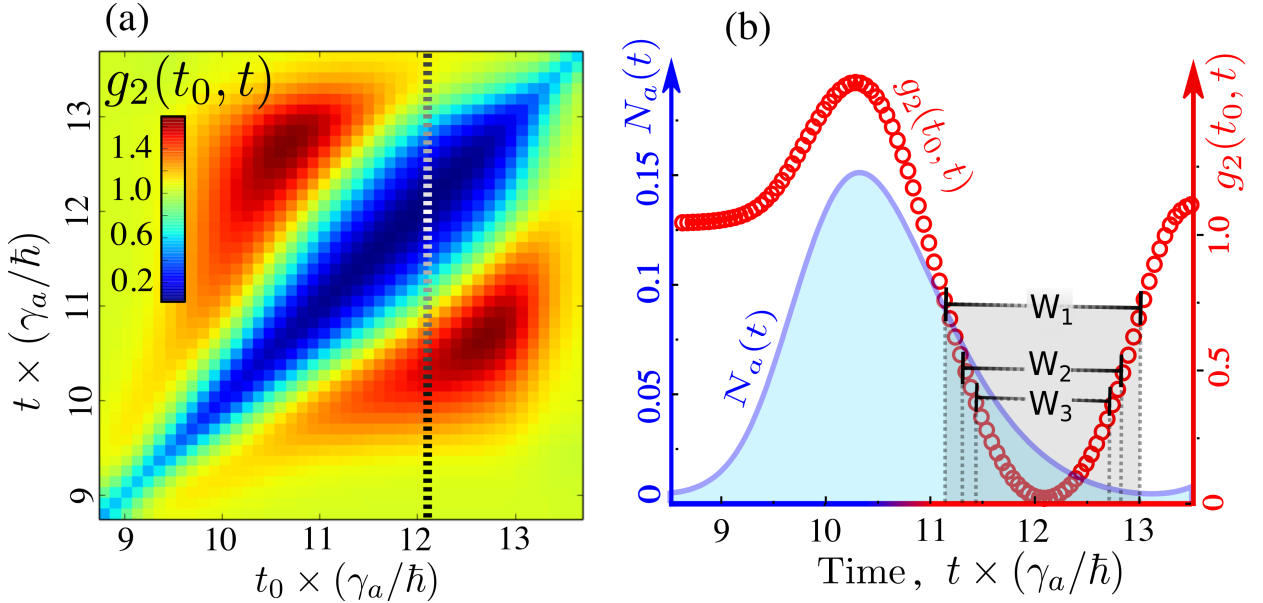}
\caption{(a) Color plot of the unequal time correlation function $g_2(t_0,t)$ as a function of $t_0$ and $t$ in the span of one incoherent pulse. We find a fish-shaped region (blue area) where the probability ($g_2(t_0,t)$) of two sequential emissions of photons is low. In each pulse, there is an optimum $t_0$ (indicated by the vertical dotted line) maximizing the $t$ window in which $g_2(t_0,t)<1$. (b) We superimposed two plots: $N_a(t)$ and $g_2(t_0,t)$ as functions of time $t$ during one incoherent pulse with an optimum $t_0=12.08\, \hbar/\gamma_a$. We indicate 3 time windows (grey shades) corresponding to $g_2(t_0,t)<0.4$, $<0.5$ and $<0.75$ which are having widths $W_1=1.34\, \hbar/\gamma_a$, $W_2=1.51\, \hbar/\gamma_a$ and $W_3=1.89\, \hbar/\gamma_a$ respectively.}
\label{UnequalTimeG2}
\end{figure}

For our chosen parameters mentioned in Fig.~\ref{PulseTrainFig}, the antibunched population in $\ah$ mode reaches up to $N_a=0.11$ in each pulse. Thus, roughly one in every $10$ pulses will generate a single photon. Although a low $g_2(t,t)$ ensures no simultaneous multiphoton emission, it does not reveal the time gap between two consecutive emissions. For this, we compute the unequal time correlation function $g_2(t_0,t)$ where $t_0$ is a reference time. Note that as our system has no time translational symmetry, the correlation function $g_2(t_0,t)$ depends on both the arguments $t_0$ and $t$ individually. In Fig.~\ref{UnequalTimeG2}(a), we show the color plot of $g_2(t_0,t)$ within the span of one incoherent pulse. We find a fish-shaped region (blue area in the figure) in the $t_0$-$t$ plane where $g_2(t_0,t)$ is small. This fish-shaped region corresponds to low probability of consecutive emissions. The width of the region along $t$ signifies the average time gap between two consecutive emissions. We maximize this width by an optimum choice of the reference time $t_0$. For the pulse considered in Fig.~\ref{UnequalTimeG2}(a), the best value of the reference time is found to be $t_0= 3.845T_0$ i.e. $0.845T_0$ far from the previous closest pulse. By superimposing the population $N_a(t)$ and the correlation function $g_2(t_0,t)$ in Fig.~\ref{UnequalTimeG2}(b), we find the best time window where the antibunched photons has a significant population with a low $g_2(t_0,t)$. In the figure, we show 3 windows with widths $W_1=1.34\, \hbar/\gamma_a$, $W_2=1.51\, \hbar/\gamma_a$ and $W_3=1.89\, \hbar/\gamma_a$ and centered at $t_0$ corresponding to $g_2(t_0,t)<0.4$, $<0.5$ and $<0.75$ respectively. Within these time windows, the maximum values of $N_a$ varies in between $0.09$ to $0.065$. Thus, if these time windows are chosen for single photon emission, we can get one antibunched photon in every $12$ to $16$ pulses. Note that $t_0$ is chosen $3.845T_0$ for the considered pulse in Fig.~\ref{UnequalTimeG2}(b), but for subsequent pulses $t_0=(n+3.845)T_0$ where $n$ is an integer.

{\bf Conclusion.---} We presented the general idea that a nonlinear mode weakly coupled to a mode exhibiting gain can be utilized to produce antibunched photons. When an incoherent excitation is applied, with a rate smaller than the dissipation rate of the gain mode, the system attains a strongly antibunched steady state ($g_2 \sim 0$). We investigated the steady state properties of the system both analytically and numerically by solving the quantum master equation for an applied incoherent pump. The achieved analytical solutions for the photon populations in both modes agree exactly with the numerically calculated results. We further derived the equal time second order correlation function analytically, which also agrees well with numerical values in the most relevant parameter range. We found that the performance of the single photon source is optimum when the mode coupling $J$ and the nonlinearity in the pumped mode are weak, but the nonlinearity in the antibunched mode must be strong.

In the case of pulsed gain (or incoherent excitation), the nonlinear mode shows strong antibunching only when the photon population is significant in the mode. Thus, the system can be used as a probabilistic source of single photons triggered at specific times. We calculated the unequal time second correlation function during the span of a pulse, and found that the single photon emission would be well separated in time with a gap comparable to the pulse period.

We identify exciton-polaritons in semiconductor microcavities as a promising platform for realization of the proposal. Recent experimental reports showed  a weak polariton blockade under coherent excitation~\cite{Matutano17} and separate experiments have reached the strongly nonlinear regime~\cite{Sun17,Rosenberg18,Togan18}. The gain medium could be realized with optically or electrical injection techniques, that is, polariton lasers operating below threshold could be used as compact probabilistic quantum sources.

Finally, we note that a significant amount of physics has been uncovered related to the blockade physics of two coupled quantum modes under coherent drive, including the influence of polarization~\cite{Bamba11APL}, the control allowed by multiple sources~\cite{Xu14,Shen15,Shen18}, antibunching of symmetric and antisymmetric modes~\cite{Xu14Sep}, and different forms of nonlinear interaction~\cite{Majumdar13,Gerace14,Zhou15,Zhou16}. It would be interesting to see the influence of similar effects in the case of a gain medium and the generalization of applications based on the photon blockade such as quantum diodes\cite{Mascarenhas14,Shen14}.

\section*{Acknowledgement}
This work was supported by the Ministry of Education (Singapore), grant MOE2017-T2-1-001.

\bibliography{references}

\begin{thebibliography}{48}
\bibitem [{\citenamefont {Footnote-1}()}]{FootNote2}%
  \BibitemOpen
  \bibfield  {author} {\bibinfo {author} {\bibnamefont {Footnote-1}},\
  }\href@noop {} {}\bibinfo {note} {While the truncation of the Fock basis is
  itself an approximation, its accuracy is straightforward to characterize. In
  principle, arbitrary accuracy can be achieved, while in practice we refer to
  the calculation as exact when accuracy to the relevant plot resolution is
  reached.}\BibitemShut {Stop}%
\end{thebibliography}

\end{document}